# AIS-Based Maritime GNSS RFI Monitoring With Multi-Vessel Coherence and Communication-Integrity Artifact Mitigation

**Sanghyeon Park[1], DeukJae Cho[2], Pyo-Woong Son[1]**

**[1] Department of Electronics Engineering, Chungbuk National University, Cheongju, 28644, Republic of Korea**
**[2] Maritime Digital Transformation Research Center, Korea Research Institute of Ships and Ocean Engineering, 34103, Daejeon, Republic of Korea**

**Correspondence**
Pyo-Woong Son. E-mail: pwson@cbnu.ac.kr

**Abstract**
Global Navigation Satellite System (GNSS) radio-frequency interference (RFI), including spoofing and jamming, threatens maritime positioning, navigation, and timing. Automatic Identification System (AIS) reports provide GNSS-derived vessel positions, offering an opportunistic wide-area monitoring source; however, raw AIS streams contain communication-integrity artifacts, such as duplicated ship identifiers and stale-data retransmissions, that can mimic trajectory distortion or reporting outages, and isolated single-vessel anomalies cannot be unambiguously attributed to GNSS interference. This paper proposes an AIS-based framework that mitigates communication-integrity artifacts, generates kinematic-consistency and transmission-continuity anomaly cues, and groups retained cues into multi-vessel events using spatiotemporal Density-Based Spatial Clustering of Applications with Noise (ST-DBSCAN). Applied to approximately 966 million AIS messages from Korean coastal waters, the framework retained 48 spoofing-consistent distortion events and 98 jamming-consistent reporting-outage events. These results suggest that AIS can serve as a complementary source for maritime GNSS RFI monitoring.

## 1 INTRODUCTION

Maritime navigation and surveillance systems rely critically on the integrity of Positioning, Navigation, and Timing (PNT) information (International Civil Aviation Organization [ICAO], International Maritime Organization [IMO], & International Telecommunication Union [ITU], 2025). The Automatic Identification System (AIS), mandated by the International Maritime Organization for safety-of-life-at-sea (SOLAS) vessels, has become a cornerstone of maritime situational awareness by continuously broadcasting vessel position, speed, and course (IMO, 2002). These AIS reports support collision avoidance, vessel traffic services, and large-scale maritime monitoring. However, because AIS position information is predominantly derived from unencrypted Global Navigation Satellite System (GNSS) signals, it is inherently vulnerable to both intentional and unintentional disruptions (International Association of Independent Tanker Owners [INTERTANKO], 2019).

In recent years, multiple incidents have demonstrated the operational impact of GNSS degradation in maritime environments. Large-scale GNSS spoofing and jamming events have been documented across the Baltic Sea, Eastern Mediterranean, and Black Sea regions, affecting both aviation and maritime operations (Felux et al., 2024; European Union Aviation Safety Agency [EASA], 2023). In 2023–2024, spoofing incidents displaced over 100 vessels simultaneously to false airport positions in the Eastern Mediterranean (Lo et al., 2025), while AIS manipulation for sanctions evasion has been increasingly reported (Androjna et al., 2024).

GNSS spoofing can distort vessel trajectories by manipulating position solutions, while GNSS jamming can completely disrupt navigation by preventing receivers from obtaining valid fixes (IMO, 2021). Such events pose serious risks to maritime safety, particularly in congested coastal waters and strategic sea lanes. Consequently, there is growing interest in AIS-based methods for detecting GNSS interference at regional scales, where dedicated GNSS monitoring infrastructure may be sparse or unavailable (International Association of Marine Aids to Navigation and Lighthouse Authorities [IALA], 2016).

A substantial body of research has explored AIS-based anomaly detection using data-driven trajectory analysis, physical kinematic modeling, or hybrid approaches. A comprehensive review by Wolsing et al. (2022) surveyed 44 research articles on AIS track anomaly detection, identifying ten groups of detection methods ranging from clustering to deep learning. Data-driven methods identify statistical deviations from historical traffic patterns, while kinematic models evaluate whether reported vessel motion violates physical constraints.

Although these approaches can detect anomalous vessel behavior, they face two fundamental challenges when applied to GNSS interference monitoring. First, AIS data are affected by a wide range of non-GNSS-related irregularities, including communication-layer faults and onboard sensor issues, which can easily be misclassified as interference-induced anomalies (El Mekkaoui et al., 2022; Harati-Mokhtari et al., 2007). Second, many existing methods analyze vessels independently and therefore fail to capture multi-vessel coherence, a defining characteristic of wide-area GNSS spoofing and jamming (Ferreira et al., 2022). Statistical frameworks for detecting malicious AIS spoofing have been proposed (d'Afflisio et al., 2021), yet these approaches still lack explicit mechanisms to distinguish communication-layer artifacts from genuine interference.

AIS anomalies do not necessarily imply GNSS interference. Communication-layer artifacts, including Maritime Mobile Service Identity (MMSI) duplication and delayed retransmission, may generate artificial trajectory distortions, while vessel-specific sensor issues such as degraded GNSS reception or transient measurement faults can produce kinematic inconsistencies similar to spoofing-induced behavior. Without explicitly accounting for these effects, AIS-based anomaly detection may incorrectly interpret non-interference irregularities as GNSS-related events.

Another critical limitation of prior studies is their treatment of GNSS jamming. Unlike spoofing, which manifests as distorted trajectories, jamming often results in the absence of valid position reports altogether (INTERTANKO, 2019). During such outages, AIS messages may cease or be transmitted at irregular intervals, rendering conventional trajectory-based detection ineffective (ITU, 2014). Robust detection of jamming therefore requires explicit analysis of transmission continuity in addition to kinematic behavior (Last et al., 2014).

In this context, AIS-based GNSS RFI monitoring should be framed as a multi-vessel coherence-based inference problem rather than as a deterministic classifier for isolated vessel anomalies. A single-vessel AIS irregularity cannot be uniquely attributed to GNSS interference because similar observations may result from MMSI duplication, communication artifacts, onboard sensor faults, or normal operational behavior. The proposed framework therefore uses multi-vessel spatiotemporal coherence as corroborating evidence for GNSS-RFI-consistent events, while treating isolated or vessel-confined anomalies conservatively as communication- or sensor-integrity artifacts.

To address these challenges, this paper proposes a hierarchical AIS-based framework for maritime GNSS RFI monitoring. Unlike conventional AIS anomaly-detection approaches, the proposed framework explicitly separates communication-integrity artifacts, vessel-specific sensor-integrity artifacts, and multi-vessel coherent anomaly patterns before interpreting potential GNSS-RFI-related events. The framework is evaluated using approximately 966 million AIS messages collected in Korean coastal waters, demonstrating how AIS observations can serve as a complementary source for wide-area maritime GNSS RFI monitoring.

The remainder of this paper is organized as follows. Section 2 reviews related work. Section 3 presents the proposed methodology. Section 4 describes the dataset and experimental setup. Section 5 presents the results and case studies. Section 6 discusses limitations and practical implications. Section 7 concludes the paper.

## 2 RELATED WORK

### 2.1 Communication integrity and AIS data quality issues

AIS communication itself introduces multiple data-quality issues that can mimic GNSS interference and inflate false alarm rates if not explicitly addressed (El Mekkaoui et al., 2022). A well-documented example is MMSI duplication, where distinct vessels broadcast the same identifier due to configuration errors or reused transponders (Harati-Mokhtari et al., 2007; Zhao et al., 2018). Shore-station or receiver time-synchronization errors can also produce contradictory multi-position reports for a single MMSI (Fukuda et al., 2024). Meanwhile, oscillatory jitter or retransmission of stale data commonly results in physically inconsistent trajectories that resemble spoofing-induced distortions (ITU, 2014). Additionally, infrastructure-induced inconsistencies, such as multi-receiver rebroadcast delays, may generate position scattering that confounds naive anomaly detectors.

These communication-layer artifacts originate from the AIS transmission chain rather than from navigation sensor degradation. Nevertheless, they are frequently misclassified as GNSS anomalies in existing AIS-based detection studies, underscoring the necessity of integrating communication integrity assessment into AIS-based monitoring frameworks. Without such diagnostics, anomaly detection results cannot be reliably interpreted or operationalized (Iphar et al., 2020).

### 2.2 AIS data-driven anomaly detection

A large body of AIS anomaly detection research has relied on data-driven approaches, where normal vessel behavior is modeled from historical AIS trajectories and deviations are identified statistically. Classical clustering methods such as Density-Based Spatial Clustering of Applications with Noise (DBSCAN) have been used to distinguish between stationary and moving clusters and to detect outliers that deviate from typical spatial patterns (Ferreira et al., 2022; Goodarzi & Shaabani, 2019; Han et al., 2021; Newaliya & Singh, 2024; Rong et al., 2020). Complementary approaches estimate expected trajectories through kinematic interpolation or prediction models and classify large residuals as anomalies (Guo et al., 2021). More recently, deep-learning-based frameworks, ranging from probabilistic neural representations to convolutional architectures, have demonstrated improved capability in capturing complex motion dependencies and detecting abnormal behavior in data-rich environments (Liang et al., 2024; Lv et al., 2025; Nguyen et al., 2022).

Despite these advancements, data-driven methods remain fundamentally constrained by the distributions on which they are trained. Their performance deteriorates when exposed to rare, emerging, or region-specific motion patterns that differ from the training set. Moreover, the anomaly decisions produced by deep-learning models often lack interpretability, providing little insight into whether the irregularity originates from GNSS degradation, vessel-specific sensor-integrity issues, or AIS communication faults. Finally, most data-driven studies focus exclusively on single-vessel trajectories, leaving them unable to detect collective anomalies, which are a defining signature of wide-area GNSS interference.

### 2.3 Dynamic-state and kinematic modeling approaches

To improve interpretability and incorporate physical constraints, several studies have applied dynamic-state estimation and kinematic modeling to AIS anomaly detection. Early work demonstrated the use of Extended Kalman Filters (EKF) to track vessel trajectories and flag physically implausible measurements (Liu et al., 2019; Mai Thi et al., 2024; Siegert et al., 2016). More advanced studies introduced the Interacting

Multiple Model (IMM) filter, which combines Constant Velocity (CV) and Constant Turn Rate and Velocity (CTRV) models to handle both straight-line and maneuvering behavior without requiring training data (Genovese, 2001; Tian et al., 2025; Yang et al., 2022).

Kinematic filtering provides clear interpretability, since detected anomalies correspond directly to violations of motion feasibility such as unrealistic acceleration or discontinuous headings. However, these methods still operate primarily on individual vessel trajectories, making them unsuitable for detecting region-level anomalies that affect multiple vessels simultaneously. Moreover, kinematic approaches alone cannot differentiate external GNSS distortion from vessel-specific sensor-integrity issues or AIS communication-layer artifacts, including MMSI duplication or delayed retransmission. Most critically, they cannot be applied when AIS position reports cease entirely during GNSS jamming, which eliminates the positional data required for kinematic evaluation.

### 2.4 Spatiotemporal clustering for maritime anomalies

Recognizing that wide-area GNSS interference often manifests as coordinated anomalies across multiple vessels, recent studies have adopted spatiotemporal clustering techniques, particularly ST-DBSCAN, to identify groups of irregular observations (Birant & Kut, 2007). ST-DBSCAN extends DBSCAN by incorporating a temporal radius in addition to a spatial radius, enabling detection of dense anomaly clusters that persist across space and time. This approach has been used to reveal abnormal regional behaviors, including both localized disruptions and broader traffic inconsistencies (Olesen et al., 2023).

However, prior applications of ST-DBSCAN typically rely on raw AIS observations or simple trajectory deviations as inputs, without incorporating domain-specific validation of kinematic feasibility or communication integrity. As a result, spatiotemporal clusters can arise from fundamentally different sources (including GNSS spoofing, MMSI duplication, multi-station synchronization errors, or sensor-integrity artifacts), yet appear indistinguishable in cluster outputs. Without upstream diagnostics or multi-layer validation, spatiotemporal clustering alone cannot reliably separate true GNSS interference from AIS system artifacts. The proposed framework addresses this gap by applying communication-integrity mitigation and vessel-level sensor-artifact separation before spatiotemporal clustering, rather than clustering raw AIS deviations directly.

## 3 METHODOLOGY

### 3.1 Overview of the unified processing pipeline

This study proposes a unified AIS-based processing pipeline designed to support maritime GNSS RFI monitoring by separating communication-integrity artifacts, vessel-specific sensor-integrity artifacts, and regional multi-vessel anomaly patterns. As illustrated in Figure 1, the pipeline consists of three hierarchical integrity-assessment stages: communication integrity assessment, sensor integrity assessment, and spatiotemporal integrity assessment.

In the communication integrity assessment stage (Stage 1), AIS messages are screened to identify and remove artifacts arising from MMSI duplication or stale-data retransmission. In the sensor integrity assessment stage (Stage 2), per-vessel anomaly cues are generated from AIS trajectories and transmission records without premature classification or removal; these cues capture heterogeneous positioning and reporting irregularities, including kinematic inconsistencies and reporting outages. In the spatiotemporal integrity assessment stage (Stage 3), the preserved cues are jointly analyzed through multi-vessel spatiotemporal clustering to differentiate vessel-specific sensor-integrity artifacts from regional multi-vessel anomaly candidates consistent with GNSS RFI.

A key design principle of the proposed framework is that only communication-integrity artifacts are removed before clustering, while other anomaly cues are retained and classified based on persistence and multi-vessel coherence. This design reduces the risk that sensor-integrity artifacts are misinterpreted as regional GNSS-RFI-consistent events.

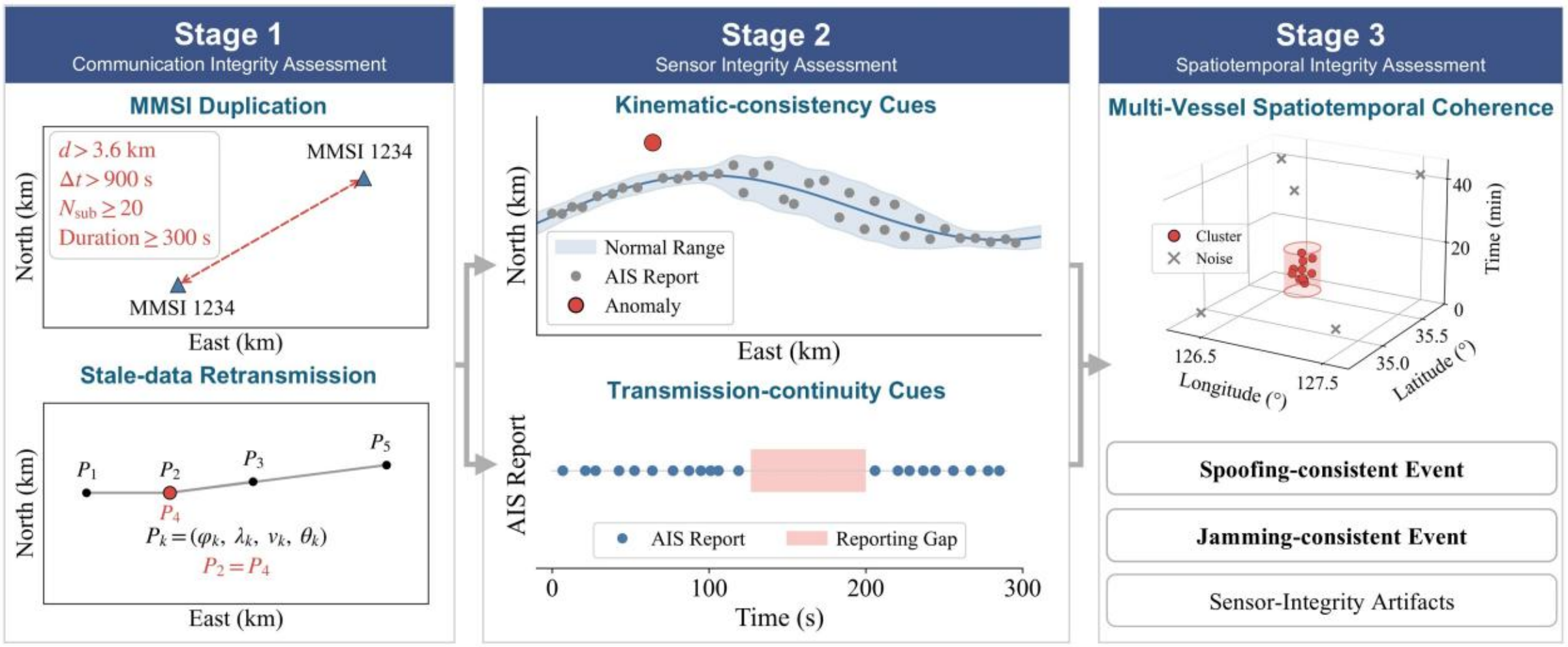

**FIGURE 1** Flowchart of the proposed hierarchical processing pipeline for AIS-based GNSS RFI monitoring, comprising communication integrity assessment, sensor integrity assessment, and spatiotemporal integrity assessment

### 3.2 Communication integrity assessment

Certain AIS anomalies originate from communication-layer faults rather than from navigation sensor degradation or GNSS interference. To prevent such artifacts from inflating false alarm rates, two representative communication-integrity issues are explicitly diagnosed and removed at an early stage: concurrent MMSI duplication and stale-data retransmission of AIS messages.

MMSI duplication occurs when multiple vessels broadcast the same identifier within overlapping time intervals while being separated by distances exceeding feasible vessel motion constraints. This artifact produces artificial position jumps in reconstructed trajectories, despite each vessel individually exhibiting physically plausible dynamics (Goodarzi & Shaabani, 2019; Olesen et al., 2023; Zhang & Li, 2022). MMSI duplication is identified using spatial and temporal consistency checks, as summarized in Algorithm 1.

**ALGORITHM 1** MMSI co-occurrence-based duplication verification

**Input:** AIS records grouped by MMSI; for each MMSI $i$, $M_i = \{(t_k, \varphi_k, \lambda_k)\}$ is sorted by time

**Params:** $\delta_d$ (spatial gap), $\delta_t$ (temporal gap), $N_{sub}$ (min points), $\tau_{sub}$ (min duration), $\varepsilon_s$, $\varepsilon_t$ (co-occurrence radii), minimum distinct MMSIs for co-occurrence gate = 2

**Definitions:** a sub-track is a maximal set of same-MMSI records connected under distgeo $\leq \delta_d$ and $|\Delta t| \leq \delta_t$. TimeRange($S$) = [$t_{start}(S)$, $t_{end}(S)$].

**Step 1: Sub-track formation and validity filtering**

for each MMSI track $M_i$ do
    partition $M_i$ into sub-tracks as maximal connected components under the neighbor relation and merge interleaved fragments
    keep valid sub-tracks with $|S_j| \geq N_{sub}$ and duration($S_j$) $\geq \tau_{sub}$
end for

**Step 2: Per-MMSI overlap-candidate collection**

$E \leftarrow \emptyset$
for each MMSI with two or more valid sub-tracks having overlapping TimeRange do
    add $M_i$ to $E$ with representative points (sub-track centroids and time ranges)
end for

**Step 3: Cross-MMSI co-occurrence gate**

form spatiotemporal components of $E$ under distgeo $\leq \varepsilon_s$ and time-range gap $\leq \varepsilon_t$
for each component $C$ do
    if $C$ contains $\geq 2$ distinct MMSIs then

retain all involved MMSIs as anomaly candidates
else
label the isolated MMSI's overlapping records as an MMSI-duplication artifact and exclude
end if
end for
**Output:** record labels (MMSI-duplication artifact / retained anomaly candidate)

Algorithm 1 operationalizes this co-occurrence test as follows: an MMSI with two or more temporally overlapping sub-tracks is flagged as an overlap candidate, and the overlap candidates are grouped into spatiotemporal components using the same radii ($\varepsilon_s$ and $\varepsilon_t$) as the event clustering. A component containing two or more distinct MMSIs is retained for subsequent sensor- and spatiotemporal-integrity assessment, because the simultaneous appearance of the same signature across multiple distinct identifiers in one neighborhood is more consistent with an external positioning disturbance than with independent identifier duplication; an isolated overlap candidate is instead excluded as a genuine MMSI-duplication artifact. This rule reflects the intended scope of the framework. An isolated same-MMSI overlap cannot be unambiguously distinguished from either genuine MMSI duplication or a single-vessel spoofing-like displacement using AIS observations alone. Therefore, the co-occurrence gate prioritizes regional multi-vessel evidence and conservatively excludes isolated overlap cases as communication-integrity artifacts, while retaining co-occurring overlap signatures for subsequent GNSS-RFI-consistent event assessment.

Stale-data retransmission occurs when previously transmitted AIS navigation tuples are rebroadcast with updated timestamps due to buffering, synchronization, or relay errors. Such retransmissions generate duplicated positions or backward motion artifacts that resemble trajectory distortion. Stale-data retransmission is identified by detecting repeated navigation tuples with inconsistent temporal ordering, as summarized in Algorithm 2.

**ALGORITHM 2** Stale-data retransmission verification process
**Input:** AIS records of a single MMSI, $M_i = \{(t_k, \varphi_k, \lambda_k, \mathrm{SO}G_k, \mathrm{CO}G_k)\}$ sorted by time
**Params:** $V_{stale}$ (minimum Speed Over Ground (SOG) above which a repeated navigation tuple is flagged as stale)
**Definitions:**
The navigation tuple of record $k$ is $\mathrm{ke}y_k = (\varphi_k, \lambda_k, \mathrm{SO}G_k, \mathrm{CO}G_k)$. Here, SOG and COG denote the speed over ground and course over ground, respectively. $D$ is a lookup table that stores each previously observed navigation tuple together with the timestamp of its first occurrence within the same MMSI stream.
**Step 1: Initialization**
$D \leftarrow \emptyset$
**Step 2: Sequential scan with motion-aware filtering**
for each record ($t_k$, $\mathrm{ke}y_k$) in chronological order do
if $\mathrm{ke}y_k \in D$ then
if $\mathrm{SO}G_k \geq V_{stale}$ then
Label record $k$ as a stale-data retransmission artifact and exclude it from anomaly candidates
end if // $\mathrm{SO}G_k < V_{stale}$: the repeated low-speed report is retained and deferred to subsequent coherence analysis
else
$D[\mathrm{ke}y_k] \leftarrow t_k$ // record first occurrence
end if
end for
**Output:** Updated record labels (Stale-data retransmission artifact / anomaly candidate)

A related boundary case is position-freezing or replay-style behavior, which may produce repeated navigation reports. In shipborne AIS installations, position, SOG, and COG are normally provided to the AIS from GNSS sensor outputs through the dynamic-data interface (IMO, 2003). Therefore, the proposed stale-data retransmission diagnostic is intentionally motion-aware rather than position-only. A repeated position report is not removed unless the complete navigation tuple (φ, λ, SOG, COG) is repeated and the vessel is reported to be in motion, that is, with a reported SOG at or above the stale-speed threshold $V_{stale}$. This design preserves low-speed or stationary repeated reports, including possible position-freezing behavior, for subsequent multi-vessel coherence analysis instead of prematurely removing them as stale-data retransmission. Conversely, complete navigation-state repetition while the vessel is reported to be in transit is more consistent with AIS buffering, relay aggregation, decoder duplication, or stale-data retransmission than with a newly generated GNSS navigation solution. Such records are therefore treated as communication-integrity artifacts rather than as GNSS-RFI-consistent anomaly cues.

Because both artifacts originate from the AIS communication chain rather than from vessel navigation dynamics or onboard GNSS performance, records labeled as communication-integrity artifacts are removed from further analysis once identified. No kinematic-consistency or sensor-integrity anomaly cues are removed at this stage, ensuring that subsequent classification is applied to retained anomaly cues after communication-integrity artifacts have been mitigated.

### 3.3 Sensor integrity assessment

After communication integrity assessment, the remaining AIS data may still contain irregularities originating from navigation sensor degradation or external GNSS interference. Rather than attempting to classify or remove such anomalies in isolation, the proposed framework extracts anomaly cues that are preserved for subsequent spatiotemporal clustering. Two complementary types of anomaly cues are generated: kinematic-consistency cues derived from vessel motion and transmission-continuity cues derived from AIS reporting behavior.

#### 3.3.1 Kinematic-consistency anomaly cues via IMM

For AIS trajectories with valid position reports, kinematic-consistency cues capture distortion-like behavior using an IMM filter combining CV and CTRV motion models as depicted in Figure 2. The IMM filter estimates vessel state while accommodating both straight-line and maneuvering behavior without requiring training data.

The vessel state vector at time step k is defined as

$$x_k = \left[p_x, p_y, v, \psi, \dot{\psi}\right]^T \tag{1}$$

where $p_x$ and $p_y$ represent the planar position components (m), $v$ is the speed (m/s), $\psi$ is the heading angle (rad), and $\dot{\psi}$ is the yaw rate (rad/s).

Each motion model propagates the state through its dynamic equations with corresponding process noise covariance. After prediction, the IMM filter fuses the individual model estimates using normalized mode probabilities:

$$x_k = \textstyle\sum_i \mu_{i,k}\, x_{i,k},\; P_k = \sum_i \mu_{i,k} \left[P_{i,k} + (x_{i,k} - x_k)(x_{i,k} - x_k)^T\right] \tag{2}$$

where $x_{i,k}$ and $P_{i,k}$ are the state estimate and covariance of model $i$, $\mu_{i,k}$ is its normalized mode probability, and $x_k$ and $P_k$ are the fused state estimate and covariance matrix.

Large residuals between reported AIS measurements and IMM-predicted states indicate violations of physical motion feasibility, such as abrupt position jumps, unrealistic accelerations, or discontinuous velocity changes. These residuals are recorded as kinematic anomaly cues. Importantly, such cues may arise from either vessel-specific sensor-integrity issues or external GNSS distortion; their final attribution is therefore deferred to the spatiotemporal clustering in Section 3.4, where multi-vessel coherence is assessed. In this implementation, the apparent speed $\hat{v}$, defined here as the innovation-derived value and distinct from the state-vector speed $v$ in (1), is estimated from the position jump relative to the model prediction, normalized by the inter-report interval; a kinematic-consistency anomaly cue is flagged when $\hat{v} \geq v_{\text{th}}$, where $v_{\text{th}}$ is the conservative kinematic-screening threshold listed in Table 2.

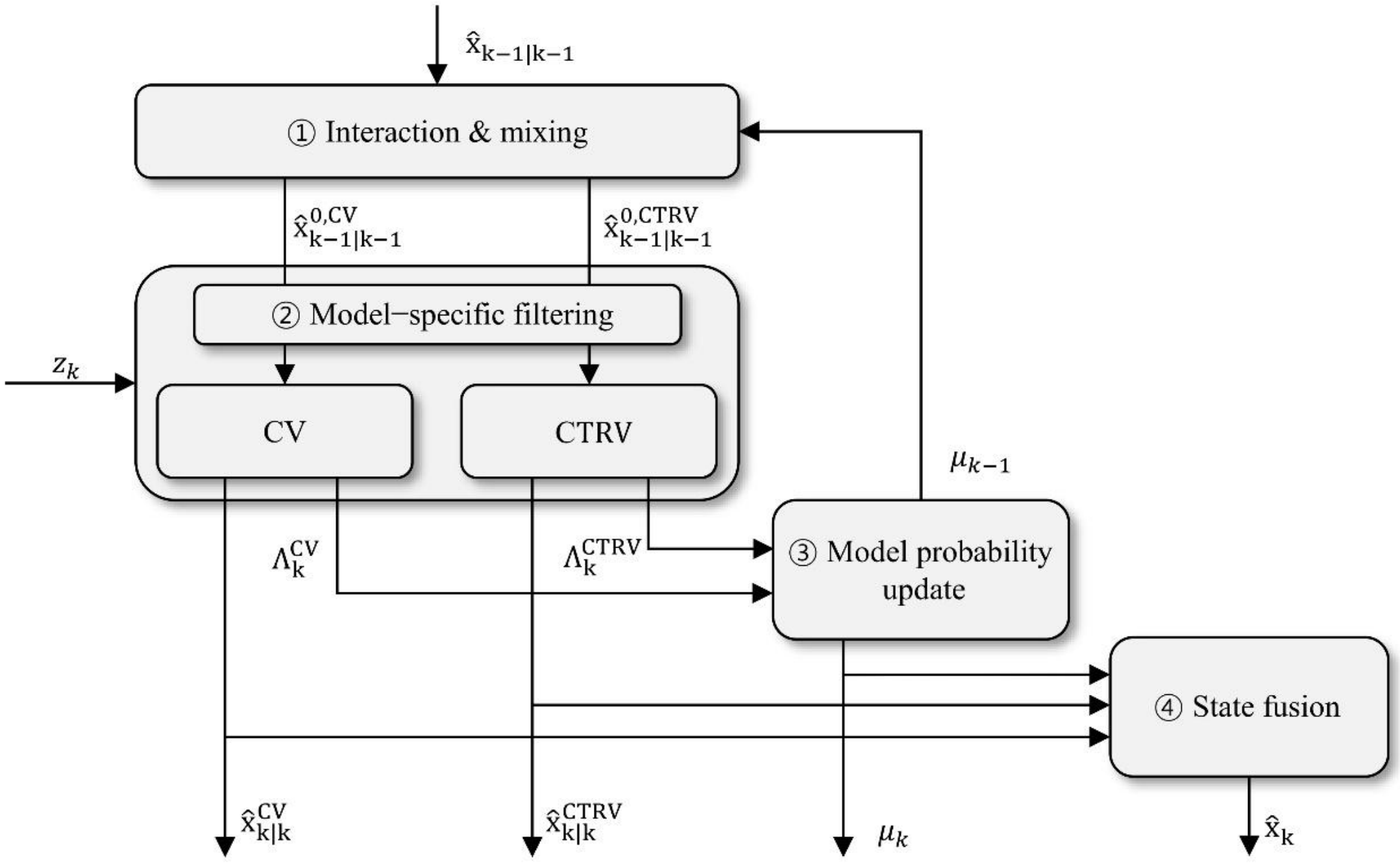

**FIGURE 2** Structure of the Interacting Multiple Model (IMM) filter combining Constant Velocity (CV) and Constant Turn Rate and Velocity (CTRV) motion models for kinematic-consistency screening of AIS trajectories

### 3.3.2 Transmission-continuity anomaly cues

Whereas the kinematic-consistency cues in Section 3.3.1 target distortion-like behavior, transmission-continuity cues target outage-like behavior, in which a GNSS receiver that fails to produce a valid position solution causes AIS position reports to cease or be transmitted at irregular intervals. These cues are extracted by analyzing inter-message intervals for each vessel. Reporting gaps exceeding the vessel-specific reporting pattern are therefore treated as outage-related anomaly cues. To capture this effect, the framework computes $\tilde{T}$, the vessel-specific median of the inter-message intervals during normal navigation (SOG ≥ $V_{stale}$), and defines a transmission-continuity threshold:

$$T_{\mathrm{th}} = max(T_{min}, \kappa\tilde{T}) \tag{3}$$

where $T_{min}$ provides a minimum threshold floor to suppress routine short reporting fluctuations, and κ is the median-gap multiplier. An inter-message interval $T_k$, defined as the elapsed time between two consecutive reports of the same vessel, is marked as a transmission-continuity anomaly cue if:

$$T_k > T_{\mathrm{th}}. \tag{4}$$

These transmission-continuity anomaly cues are preserved without classification, as isolated outages may reflect vessel-specific receiver issues, whereas simultaneous outages across multiple vessels may indicate a regional reporting-outage pattern consistent with GNSS RFI. Final interpretation is deferred to the clustering stage, where multi-vessel coherence can be assessed.

### 3.4 Spatiotemporal integrity assessment

In the final stage of the unified processing pipeline defined in Section 3.1, all preserved anomaly cues are embedded into a joint spatiotemporal space and analyzed using the ST-DBSCAN algorithm. By extending DBSCAN with temporal constraints, ST-DBSCAN enables systematic grouping of anomalies that are coherent in both space and time, thereby providing the basis for final anomaly categorization. For each anomaly candidate $i$, the spatial and temporal separations from another candidate $j$ are defined as

$$d_{\mathrm{space}}(i,j) = \mathrm{dist}_{\mathrm{geo}}((\phi_i, \lambda_i), (\phi_j, \lambda_j)), d_{\mathrm{time}}(i,j) = | t_i - t_j |, \tag{5}$$

where $\text{dist}_{\text{geo}}(\cdot)$ denotes the great-circle (geodesic) distance on the World Geodetic System 1984 (WGS-84) ellipsoid, and $d_{\text{time}}(i,j)$ is the absolute time difference between the two anomaly candidates. A pair of anomalies $i$ and $j$ are regarded as neighbors when:

$$d_{\text{space}}(i,j) < \varepsilon_s \text{ and } d_{\text{time}}(i,j) < \varepsilon_t, \tag{6}$$

where $\varepsilon_s$ and $\varepsilon_t$ denote the baseline spatial and temporal radii, respectively. These values were selected empirically through a grid search over candidate spatial and temporal radii to capture localized multi-vessel anomaly patterns while avoiding excessive merging of unrelated events.

Based on these definitions, ST-DBSCAN performs anomaly grouping and final categorization through the following steps:

(1) Anomalies are first grouped according to spatiotemporal proximity, forming candidate clusters that may involve either single or multiple vessels.

(2) Cluster composition is then evaluated to distinguish isolated vessel-level anomalies from multi-vessel structures sharing common spatial and temporal bounds.

(3) The temporal persistence of each cluster is examined to differentiate sustained irregularities from short-lived events.

Through this process, single-vessel clusters exhibiting sustained extreme deviations are classified as persistent sensor-integrity artifacts, whereas short-lived isolated clusters are classified as transient sensor-integrity artifacts. In contrast, clusters involving multiple vessels are interpreted as regional GNSS-RFI-consistent anomaly candidates. When such multi-vessel clusters primarily exhibit coordinated trajectory distortion, they are classified as spoofing-consistent distortion events, whereas clusters characterized by concurrent reporting outages are classified as jamming-consistent localized reporting-outage events.

To ensure that only statistically meaningful multi-vessel structures are interpreted as wide-area events, additional vessel-level coherence criteria are applied as part of the final categorization. Specifically, a spatiotemporal cluster is regarded as a valid multi-vessel event only when it satisfies the ST-DBSCAN density condition defined by $minPts$ and the anomalous-vessel ratio within the cluster satisfies $Th_{group}$. Here, minPts denotes the minimum number of anomaly cue points required for ST-DBSCAN core formation rather than a count of distinct vessels. The final regional-event interpretation is instead made at the vessel level, requiring the involvement of multiple distinct vessels together with an anomalous-vessel ratio above the predefined threshold. These criteria prevent clusters formed by a small number of outlier vessels embedded in otherwise normal traffic from being misinterpreted as collective anomalies. Together, they ensure that the final classification reflects a statistically coherent, region-level deviation rather than incidental or vessel-specific irregularities. In addition, to distinguish localized jamming-consistent outages from receiver- or feed-level data loss, each candidate multi-vessel outage cluster is audited against the whole-feed reception during its outage interval: the cluster is retained as jamming-consistent only if AIS messages from outside the affected area continue to be received, whereas a cluster during which the entire feed also ceases is reclassified as a receiver-level outage and excluded.

## 4 DATASET AND EXPERIMENTAL SETUP

### 4.1 AIS dataset specification

The AIS dataset used in this study was obtained from the national AIS archive operated by the Korea Research Institute of Ships and Ocean Engineering (KRISO). The evaluation period spans from 1 November to 30 December 2024, covering coastal and offshore waters surrounding the Korean Peninsula within latitudes 32–37° N and longitudes 123–133° E. This evaluation period was deliberately selected because it overlaps with officially reported GNSS interference activity in Korean waters: in early November 2024, the Republic of Korea Joint Chiefs of Staff announced GPS jamming incidents affecting multiple vessels and dozens of civilian aircraft, warning ships operating in the West Sea (The Korea Times, 2024). The dataset therefore covers a period during which GNSS RFI was known to be present in the region, although such announcements do not provide the per-event timing and location detail required to label individual incidents. This region encompasses major shipping lanes, port approaches, and open-sea traffic corridors, providing a representative maritime operating environment. In total, approximately 966 million AIS messages were collected during the evaluation period, representing dense vessel traffic under both coastal and offshore conditions. These messages were transmitted by 37,460 unique vessels, which is the denominator for the vessel-level percentages reported in Table 3. The large volume and high temporal resolution of the dataset enable comprehensive assessment of spatiotemporal coherence across multiple vessels, which is essential for wide-area anomaly analysis.

AIS consists of multiple message types with different field structures. In this study, the proposed framework uses Class-A dynamic position reports, corresponding to AIS Message Types 1, 2, and 3 in ITU-R M.1371-5 (ITU, 2014). Each decoded AIS report is treated as one observation and is associated with MMSI, timestamp, WGS-84 latitude and longitude, SOG, COG, and the AIS position-accuracy indicator, as summarized in Table 1. These fields provide a homogeneous set of dynamic navigation information required for the proposed trajectory-based processing pipeline. Message types that do not consistently provide these required dynamic navigation fields, such as Message Types 17–19, were not used in this analysis. A visual sample of vessel trajectories for a single day within the evaluation period is illustrated in Figure 3, highlighting the spatial density and heterogeneity of maritime traffic in the study area. All selected AIS position reports were processed in their original temporal order without subsampling, ensuring that both short-lived and persistent anomalies could be consistently evaluated within the unified processing pipeline.

**TABLE 1** AIS fields used in the proposed framework from Class-A dynamic position reports (ITU-R M.1371-5 Message Types 1, 2, and 3)

| Field | Description | Processing role |
|---|---|---|
| MMSI | Maritime Mobile Service Identity | Track grouping and vessel association |
| Timestamp | Observation time | Temporal ordering and reporting-gap analysis |
| Latitude, Longitude | Reported geodetic position in WGS-84 | Trajectory construction and spatial anomaly detection |
| SOG | Reported speed over ground | Motion-state reference and diagnostic filtering |
| COG | Reported course over ground | Motion-state reference and diagnostic filtering |
| Position Accuracy | AIS position-accuracy indicator | Position-quality context for integrity assessment |

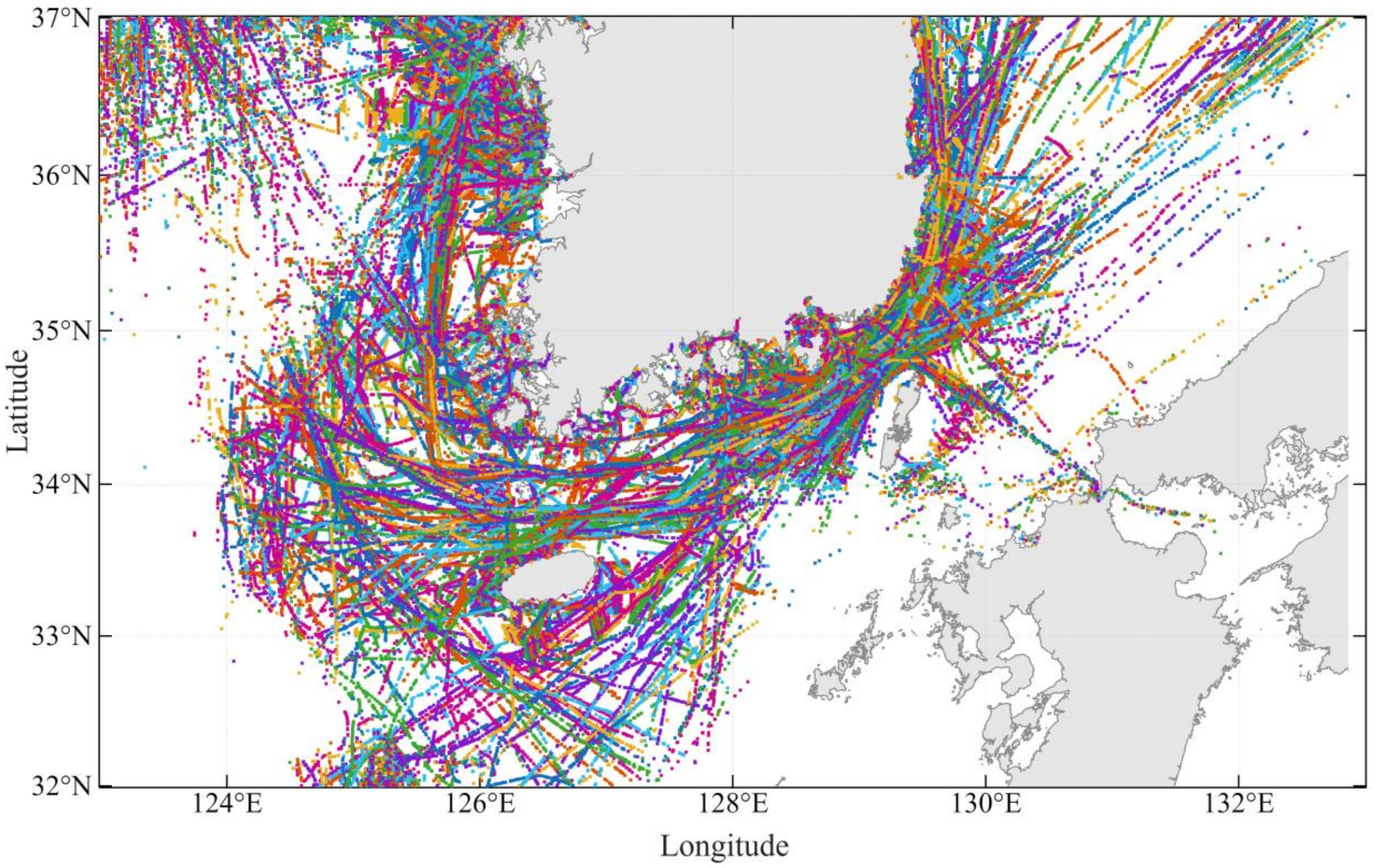

**FIGURE 3** AIS vessel trajectories on 1 November 2024 within the study area (32–37° N, 123–133° E), illustrating the spatial density and heterogeneity of maritime traffic around the Korean Peninsula

## 4.2 Data preprocessing

Prior to applying the processing pipeline described in Section 3, the AIS dataset was subjected to a limited set of preprocessing steps to remove obvious data redundancies and to constrain the analysis to the intended geographic scope. These steps were applied uniformly across the dataset and were designed to prepare the data for subsequent processing without introducing any anomaly interpretation or classification.

First, duplicate AIS records were removed (Ferreira et al., 2022; Tian et al., 2025). Records were considered duplicates when they contained identical MMSI identifiers, timestamps, and navigation tuples. This step addressed exact message replication arising from data aggregation or relay mechanisms and was performed solely to prevent redundant processing.

Second, protocol-invalid dynamic reports were excluded. AIS records outside the valid latitude and longitude ranges were removed, and records whose SOG or COG fields fell outside the valid AIS encoding range were treated as invalid decoded reports and removed.

Third, the dataset was geographically constrained by removing AIS messages located outside the vicinity of the Korean Peninsula, defined by the spatial bounds specified in Section 4.1. This filtering ensured consistency between the dataset specification and the analysis region while preventing unrelated offshore traffic from influencing the results.

No kinematic plausibility thresholds, temporal interpolation, or anomaly-specific criteria were applied during preprocessing. The SOG/COG screening described above concerned message-field validity only, not motion-based anomaly detection; therefore, no otherwise valid position report was discarded on motion-feasibility grounds.

## 4.3 Parameter configuration

Key parameter settings for each module are summarized in Table 2. Thresholds were selected to balance sensitivity to anomaly cues against robustness to routine AIS variability. For spatiotemporal clustering, the spatial radius $\varepsilon_s$ was set to reflect the typical spatial extent of local GNSS interference events in the study area, while the temporal radius $\varepsilon_t$ was set so that short-term anomaly patterns could be captured without merging unrelated incidents into a single event. For kinematic-consistency screening, the velocity limit $v_{th} = 30$ m/s was adopted as a conservative upper bound that still exceeds the design speed of most merchant vessels, thereby enforcing physical realism without penalizing legitimate high-speed navigation.

**TABLE 2** Parameter settings for each processing stage of the proposed framework

| Stage | Parameter | Value | Reference |
|---|---|---|---|
| Communication Integrity Assessment | Spatial gap for MMSI sub-track connectivity ($\delta_d$) | 3,600 m | Goodarzi & Shaabani, 2019 |
| | Temporal gap for MMSI sub-track connectivity ($\delta_t$) | 900 s (15 min) | Olesen et al., 2023 |
| | Minimum sub-track size ($N_{sub}$) | 20 AIS reports | Burger et al., 2024 |
| | Minimum sub-track duration ($\tau_{sub}$) | 300 s (5 min) | Burger et al., 2024 |
| | Minimum SOG for stale-data retransmission filtering ($V_{stale}$) | 1 knot | Paulig & Okhrin, 2024 |
| Sensor Integrity Assessment | Conservative kinematic-anomaly screening threshold ($v_{th}$) | 30 m/s | Empirical |
| | Transmission-continuity threshold based on a median-gap multiplier ($\kappa$) | 3.0 | Data-driven |
| | Transmission-continuity threshold based on a minimum threshold floor ($T_{min}$) | 60 s | Data-driven |
| Spatiotemporal Integrity Assessment | Spatial radius for spatiotemporal clustering ($\varepsilon_s$) | 10,000 m | Empirical grid search |
| | Temporal radius for spatiotemporal clustering ( $\varepsilon_t$) | 1,800 s (30 min) | Empirical grid search |
| | Minimum neighboring anomalies to form a cluster (minPts) | 5 | Hahsler et al., 2019 |
| | Anomalous-vessel ratio threshold for regional event classification ($Th_{group}$) | 60 % | Data-driven |

## 5 CASE STUDIES AND RESULTS

This section presents the results of applying the proposed hierarchical framework to the national AIS archive covering November–December 2024. Table 3 summarizes the stage-wise classification results produced by the proposed pipeline. After communication integrity assessment, sensor integrity assessment, spatiotemporal integrity assessment, and vessel-level coherence checks, the framework retained 48 spoofing-consistent distortion events and 98 jamming-consistent localized reporting-outage events. The following subsections analyze the dominant anomaly mechanisms and illustrate representative cases detected by the framework. Kinematic-consistency and transmission-continuity anomaly cues that did not satisfy the ST-DBSCAN density condition were treated as unclustered isolated cues and excluded from event-level interpretation. The Stage 2 columns in Table 3 report inclusive cue counts rather than detections: a vessel is counted once it produces any cue, so the high transmission-continuity proportion (65.19% of vessels, but only about 0.6% of position reports) reflects routine reporting gaps and is progressively constrained by clustering, multi-vessel coherence, and the whole-feed audit before any event is declared.

**TABLE 3** Stage-wise classification summary of anomaly cues across the proposed pipeline

| Analysis Stage | Process Description | Kinematic-Consistency Cues (points / vessels (%)) | Transmission-Continuity Cues (points / vessels (%)) |
|---|---|---|---|
| Stage 1: Communication Integrity Assessment | MMSI duplication artifacts removed | 976,820 points / 144 vessels (0.38%) | |
| | Stale-data retransmission artifacts removed | 2,442,344 points / 12,709 vessels (33.93%) | |
| Stage 2: Sensor Integrity Assessment | Kinematic-consistency anomaly cues | 1,008,266 points / 5,242 vessels (13.99%) | - |
| | Transmission-continuity anomaly cues | - | 5,486,796 points / 24,420 vessels (65.19%) |
| Stage 3: Spatiotemporal Integrity Assessment | Persistent sensor-integrity artifacts | 870,261 points / 76 vessels (0.20%) | - |
| | Transient sensor-integrity artifacts | 115,513 points / 2,157 vessels (5.76%) | - |
| | Spoofing-consistent distortion events | **424 points / 134 vessels (0.36%)** | - |
| | Jamming-consistent localized reporting-outage events | - | **8,480 points / 340 vessels (0.91%)** |
| | Final multi-vessel clusters | **48 events** | **98 events** |

Note: Vessel percentages are computed relative to the 37,460 unique vessels in the filtered study dataset.

### 5.1 Communication integrity assessment

A substantial portion of the initial anomalies originated not from GNSS degradation but from AIS communication-layer inconsistencies. These artifacts were systematically identified through the integrity diagnostics and removed before higher-level analysis. The patterns described here correspond to the largest contributors to false alarms in Table 3.

#### 5.1.1 Concurrent MMSI duplication

Figure 4 presents representative cases of concurrent MMSI duplication identified through the communication integrity assessment. In these cases, a single MMSI identifier is observed to broadcast AIS messages from spatially separated locations during overlapping time intervals. When these messages are naively reconstructed as a single-vessel trajectory, the resulting track exhibits abrupt long-range position jumps that violate physical continuity constraints.

However, when the messages are examined at the sub-track level, each spatially separated segment follows a smooth and dynamically consistent trajectory. The apparent anomaly therefore does not originate from navigation sensor degradation or GNSS distortion, but from identifier-level inconsistency within the AIS communication chain. Such cases are commonly associated with reused transponders, manual configuration errors, or administrative MMSI misassignment.

The proposed communication integrity assessment explicitly identifies these patterns by jointly evaluating temporal overlap and spatial separation conditions, as defined in Section 3.2. Once detected, the affected records are removed prior to sensor integrity assessment and spatiotemporal integrity assessment. As reflected in Table 3, concurrent MMSI duplication accounts for a substantial portion of the initial anomaly candidates, and its early removal prevents these artifacts from propagating into subsequent kinematic or clustering-based analyses.

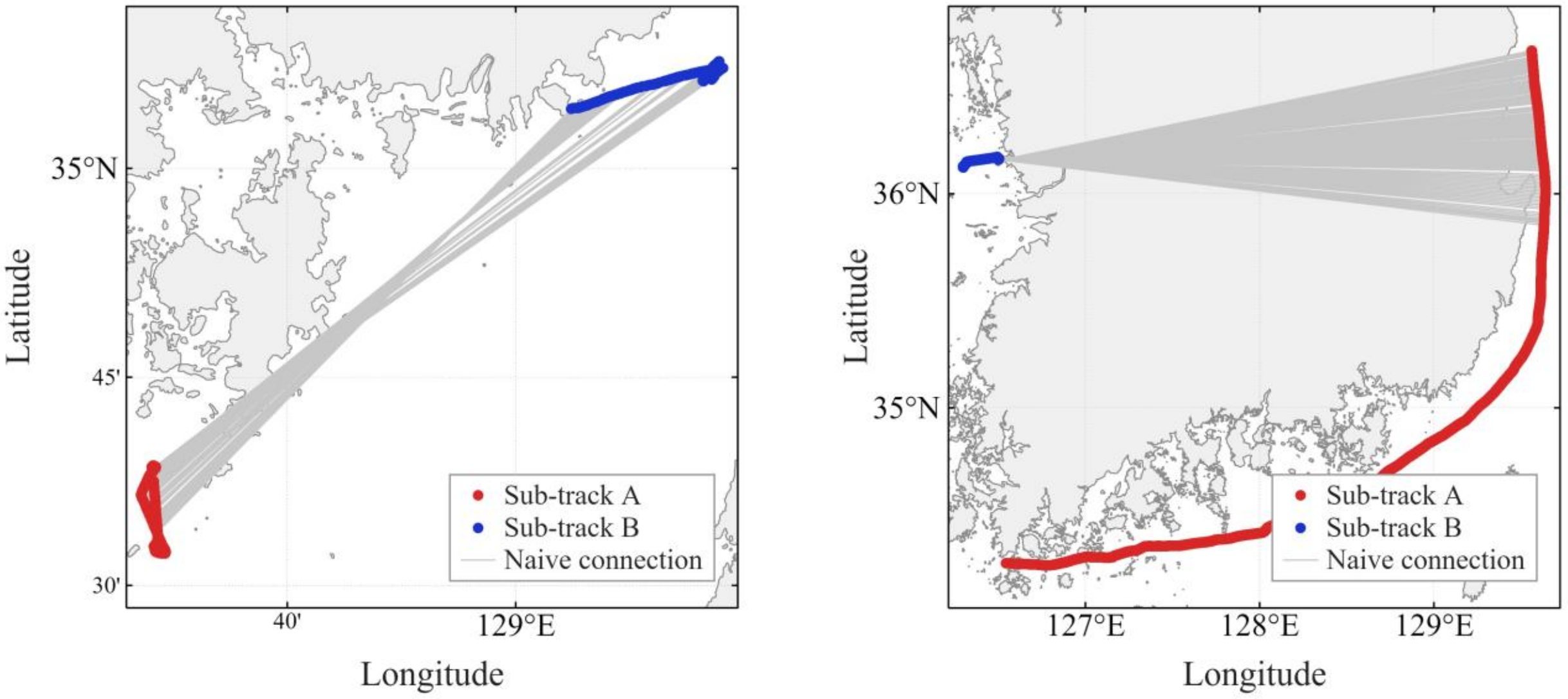

**FIGURE** 4 Representative cases of concurrent MMSI duplication. Red and blue points denote two spatially separated same-MMSI sub-tracks, while gray lines indicate infeasible naive connections that would arise if the reports were reconstructed as a single-vessel trajectory

### 5.1.2 Stale-data AIS message retransmission

Table 4 shows a representative case of stale-data retransmission, manifested as delayed AIS message rebroadcast. In this example, identical navigation tuples (comprising latitude, longitude, SOG, and COG) are observed to reappear with delayed timestamps while the vessel continues normal motion. The rebroadcast messages are separated by delays on the order of tens of seconds, despite the vessel having physically progressed along its trajectory.

When interpreted without communication-integrity checks, such delayed retransmissions generate artificial trajectory artifacts, including duplicated positions, apparent backward motion, or abrupt discontinuities. Importantly, these effects arise even though the underlying navigation data themselves remain valid. The anomaly therefore reflects irregularities in message handling, buffering, or synchronization within the AIS transmission chain rather than GNSS-induced positioning errors.

The proposed diagnostics identify stale-data retransmission by detecting repeated navigation tuples with inconsistent temporal ordering, as specified in Section 3.2. Upon identification, the affected records are excluded from further analysis. This removal step is critical for preventing retransmission-induced artifacts from being misclassified as sensor-integrity anomalies or distortion-type GNSS interference during spatiotemporal clustering. As reflected in Table 3 and illustrated by the example in Table 4, stale-data retransmissions constitute a substantial communication-level artifact category and are effectively suppressed by the proposed framework.

**TABLE 4** Example of stale-data AIS message retransmission

| Type | Timestamp | Latitude | Longitude | SOG (m/s) | COG (°) | Elapsed Time |
|---|---|---|---|---|---|---|
| Original | 16:55:17.570 | 33.046447 | 126.521270 | 7.10 | 196.2 | - |
| Rebroadcast | 16:56:14.590 | 33.046447 | 126.521270 | 7.10 | 196.2 | + 57.020 s |
| Original | 16:55:47.580 | 33.044635 | 126.520480 | 7.25 | 201.5 | - |
| Rebroadcast | 16:56:34.597 | 33.044635 | 126.520480 | 7.25 | 201.5 | + 47.017 s |

### 5.2 Sensor-integrity artifacts

#### 5.2.1 Persistent sensor-integrity artifact

Figure 5 presents representative examples of a persistent sensor-integrity artifact repeatedly observed from a single vessel throughout the entire study period. In Figure 5(a) (1 November), the vessel follows a nominal and physically plausible trajectory within its main activity area, indicating normal navigation behavior. At the same time, the track exhibits recurrent, spatially detached position scatter that accumulates into a well-defined persistent deviation area, forming a dense and geographically isolated footprint distinct from surrounding traffic. These deviations occur independently of the vessel's maneuvering state and local traffic conditions.

In Figure 5(b) (10 November), the same vessel demonstrates an analogous behavior on a different day, again maintaining a coherent nominal trajectory within the main operating region while producing intermittent but repeatedly occurring off-trajectory excursions within the persistent deviation footprint. Notably, this deviation pattern was observed on all 47 operational days during which the vessel reported AIS messages over the study period. The strict confinement of the anomaly to a single vessel, together with its long-term temporal persistence and the absence of contemporaneous deviations among nearby vessels, supports interpretation as a vessel-specific, persistent sensor-integrity artifact, rather than a shared GNSS interference event.

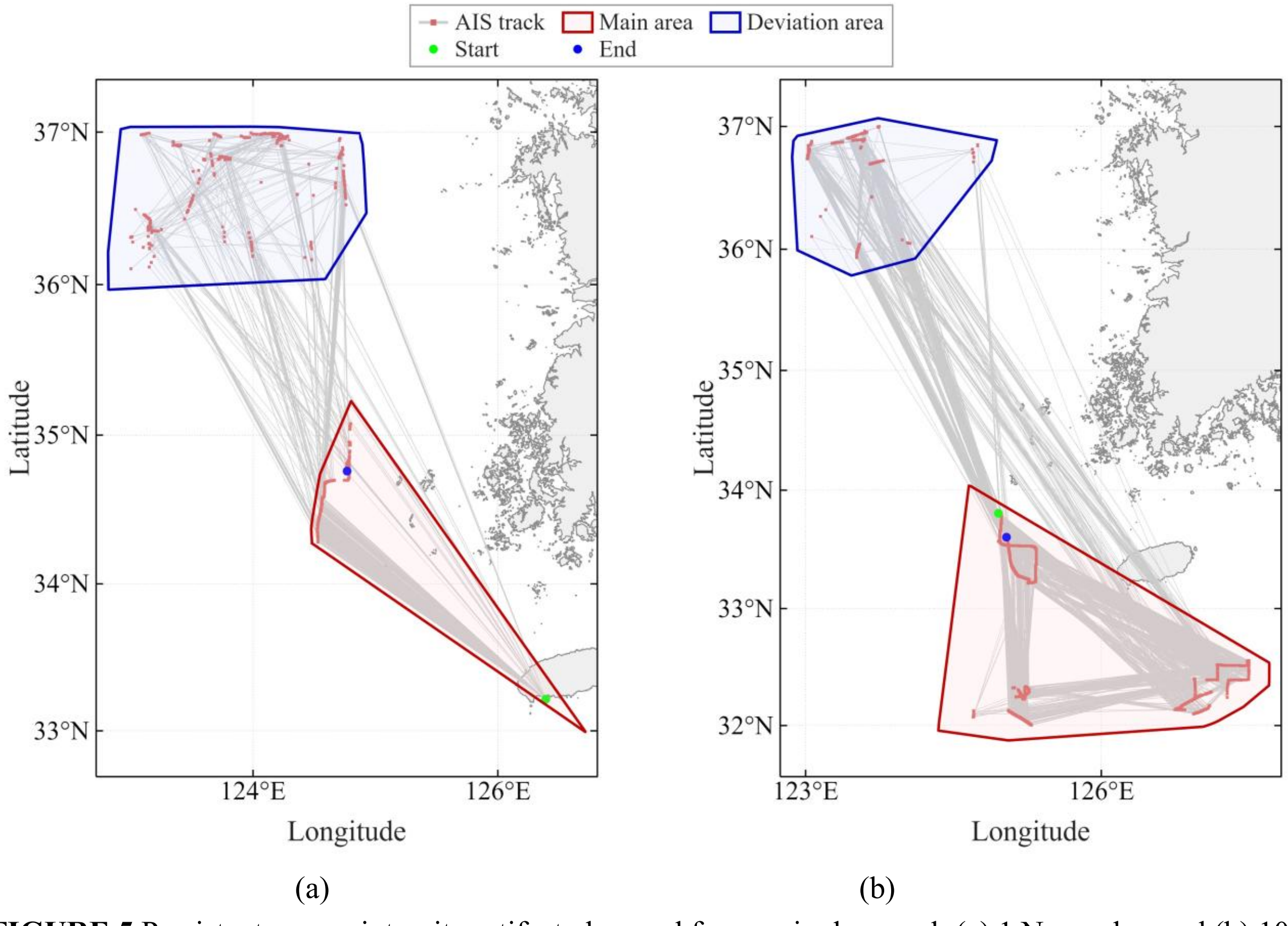

**FIGURE 5** Persistent sensor-integrity artifact observed from a single vessel: (a) 1 November and (b) 10 November 2024, showing recurrent off-trajectory excursions confined to a fixed deviation area while the vessel maintains a nominal trajectory within its main activity region

#### 5.2.2 Transient sensor-integrity artifact

Figure 6 presents representative examples of transient sensor-integrity artifacts that occur intermittently and remain confined to individual vessels. In Figure 6(a), the vessel exhibits oscillatory position deviations on six out of seventeen operational days during the study period. The anomaly displays a periodic structure that appears only under specific operating conditions, while the vessel otherwise follows a nominal trajectory. Such behavior is consistent with condition-dependent GNSS degradation or localized multipath effects, which can temporarily perturb position estimates without producing sustained or shared disturbances.

In Figure 6(b), a one-time parallel-track displacement is observed on a single day among sixteen operational days. The deviation is visually structured and, when viewed in isolation, may resemble short-lived spoofing at the trajectory level. However, the anomaly is strictly confined to a single vessel and does not recur in subsequent observations. Moreover, no contemporaneous deviations are observed among nearby vessels operating in the same spatial region.

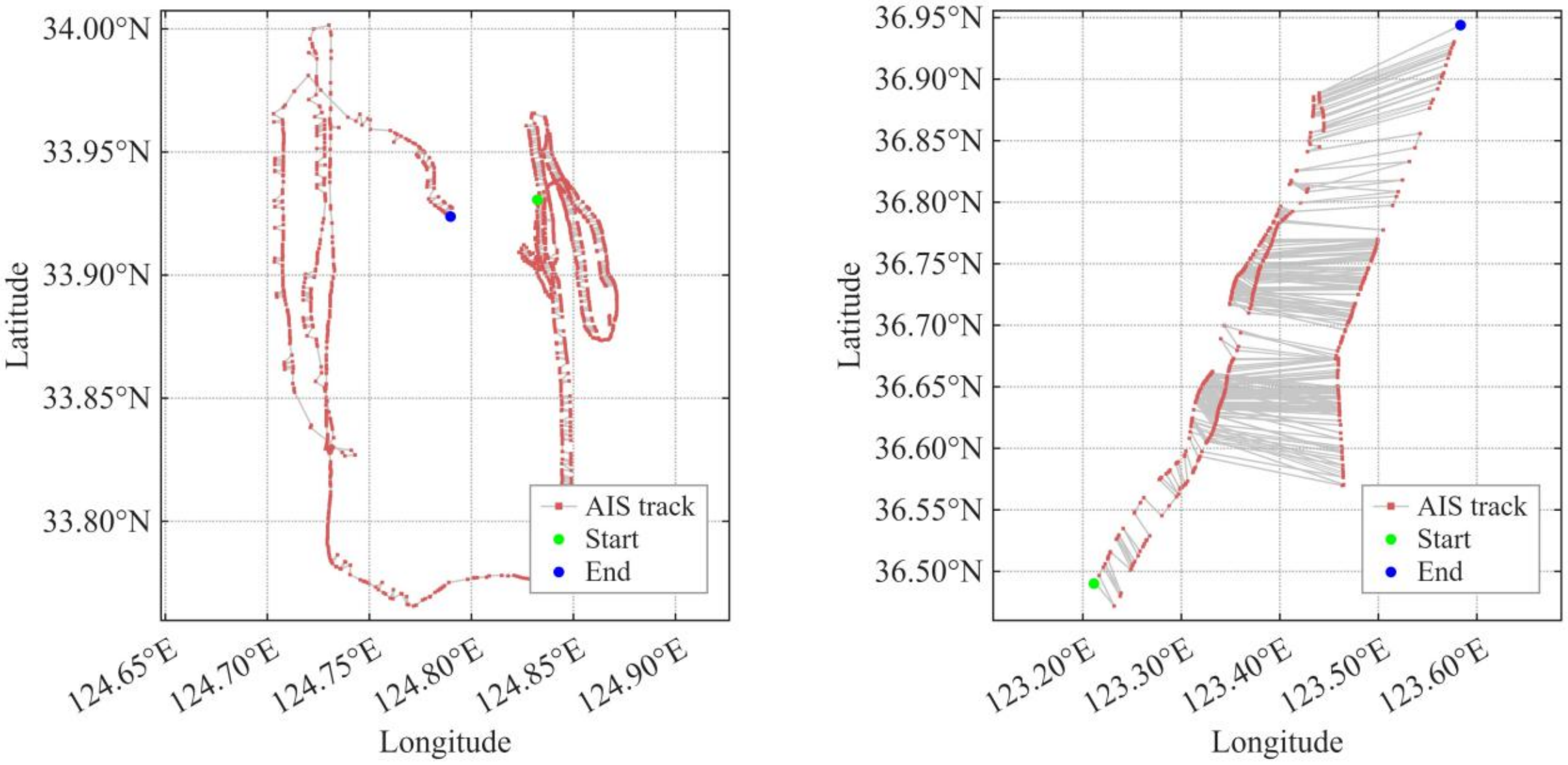

(a) (b)

**FIGURE** 6 Transient sensor-integrity artifacts confined to individual vessels: (a) oscillatory position deviations appearing on 6 of 17 operational days and (b) a one-time parallel-track displacement observed on a single day

### 5.3 Spatiotemporally shared GNSS-RFI-consistent events

#### 5.3.1 Spoofing-consistent event

Figure 7 illustrates a representative spoofing-consistent distortion event on 2 December 2024 involving two vessels, selected from the final event set. Figure 7(a) shows the spatial overview of the event within the event vicinity, and Figure 7(b) shows the corresponding per-vessel report trajectories. The reports are color-coded by elapsed time, and gray line segments connect consecutive AIS reports, highlighting repeated discontinuities between nominal trajectory segments and offset report branches while position reporting continues.

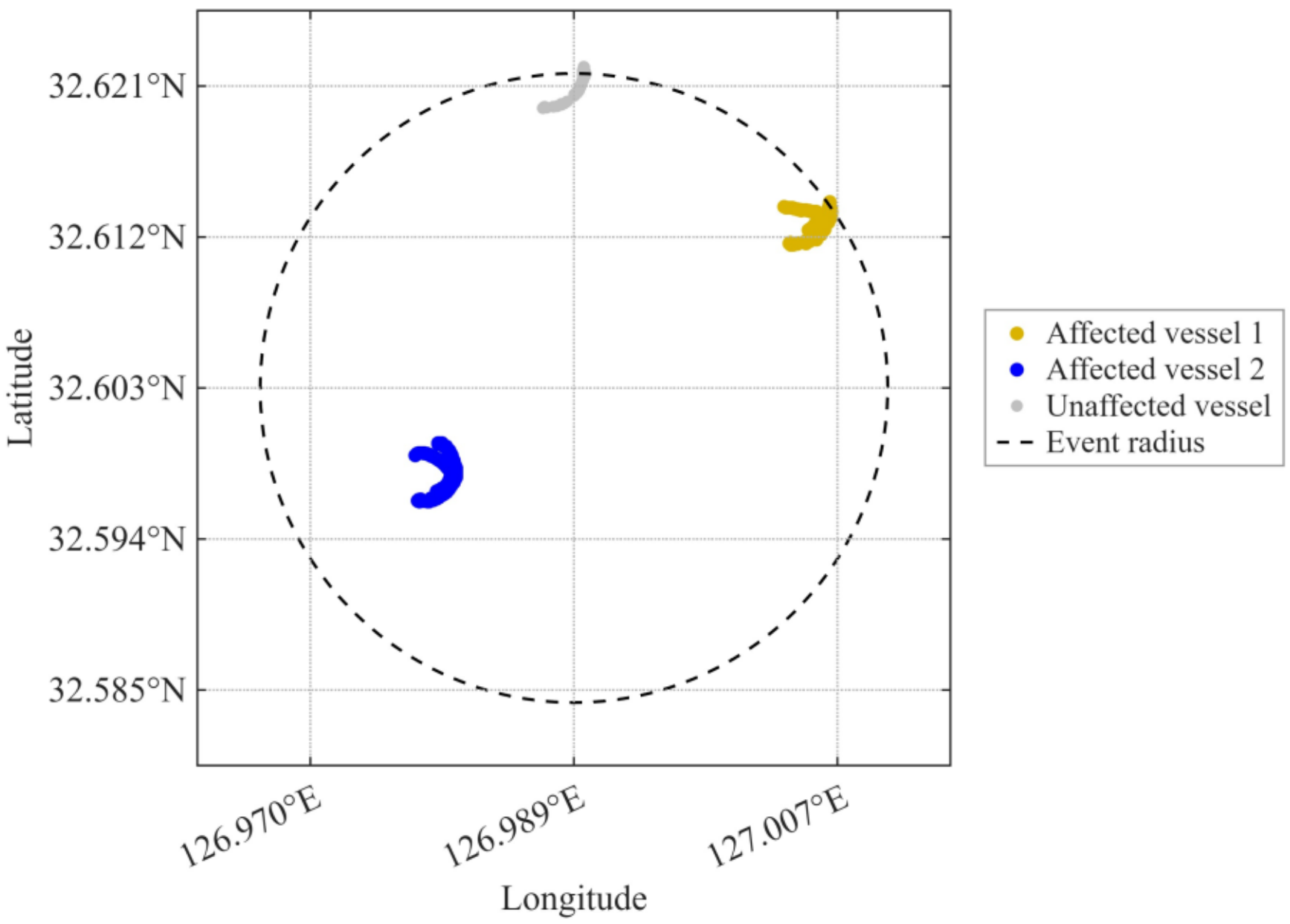

(a)

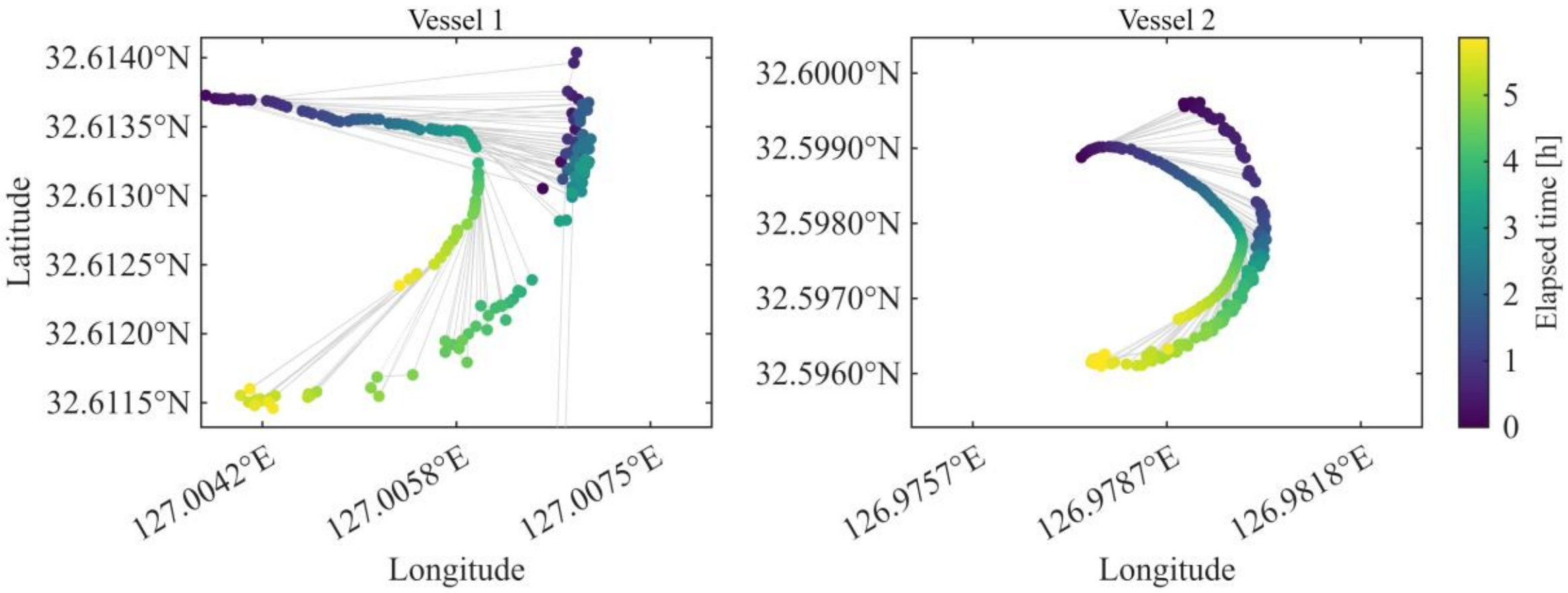

(b)

**FIGURE** 7 Representative spoofing-consistent event on 2 December 2024. (a) Spatial overview of the event, showing two affected vessels and one unaffected vessel within the event vicinity; the dashed circle indicates the approximate event radius (R ≈ 2.1 km). (b) Per-vessel report trajectories of the two affected vessels, color-coded by elapsed time; gray line segments connect consecutive AIS reports and highlight repeated discontinuities between nominal trajectory segments and offset report branches while position reporting continues

In this case, two vessels located within a radius of approximately 2.1 km exhibited recurring trajectory-discontinuity cues over an extended interval from 13:23 to 18:56 on 2 December 2024. The largest single-step jumps occurred between consecutive reports separated by approximately 1 s, with a median displacement of 197.3 m and a maximum displacement of 210.8 m. These jumps occurred while the reported SOG and COG remained nearly unchanged, indicating that the discontinuities are inconsistent with feasible vessel motion. None of the affected records corresponded to repeated navigation tuples, so the discontinuities were not attributable to the stale-data retransmission artifact described in Section 3.2.

During the event, the remaining co-located vessel within the same radius continued on a nominal trajectory without comparable discontinuities, so two of the three co-located vessels were affected. This selective impact is consistent with a localized positioning disturbance rather than a uniform communication-feed artifact. Moreover, neither of the two affected MMSIs was flagged by the MMSI-duplication verification in Algorithm 1. The two affected vessels also exhibited offset report branches of comparable scale and orientation, supporting a shared external cause while still requiring conservative interpretation because AIS data alone cannot provide signal-level confirmation.

This event was retained as a final multi-vessel cluster by the proposed framework. Although shared reception or network-level effects cannot be completely excluded from AIS data alone, the multi-vessel involvement, comparable offset-branch behavior, selective impact, and return to nominal trajectories make independent vessel-specific faults less likely. Therefore, this event is interpreted as a spoofing-consistent regional distortion event.

### 5.3.2 Jamming-consistent event

Figure 8 illustrates a representative jamming-consistent event characterized by a synchronized reporting outage across multiple vessels. In this case, five vessels located within a radius of approximately 10.0 km in open waters southwest of Jeju Island experienced a common reporting outage on 26 December 2024. Unlike distortion-type events, the anomalies manifest as abrupt interruptions in position reporting rather than as kinematic deviations.

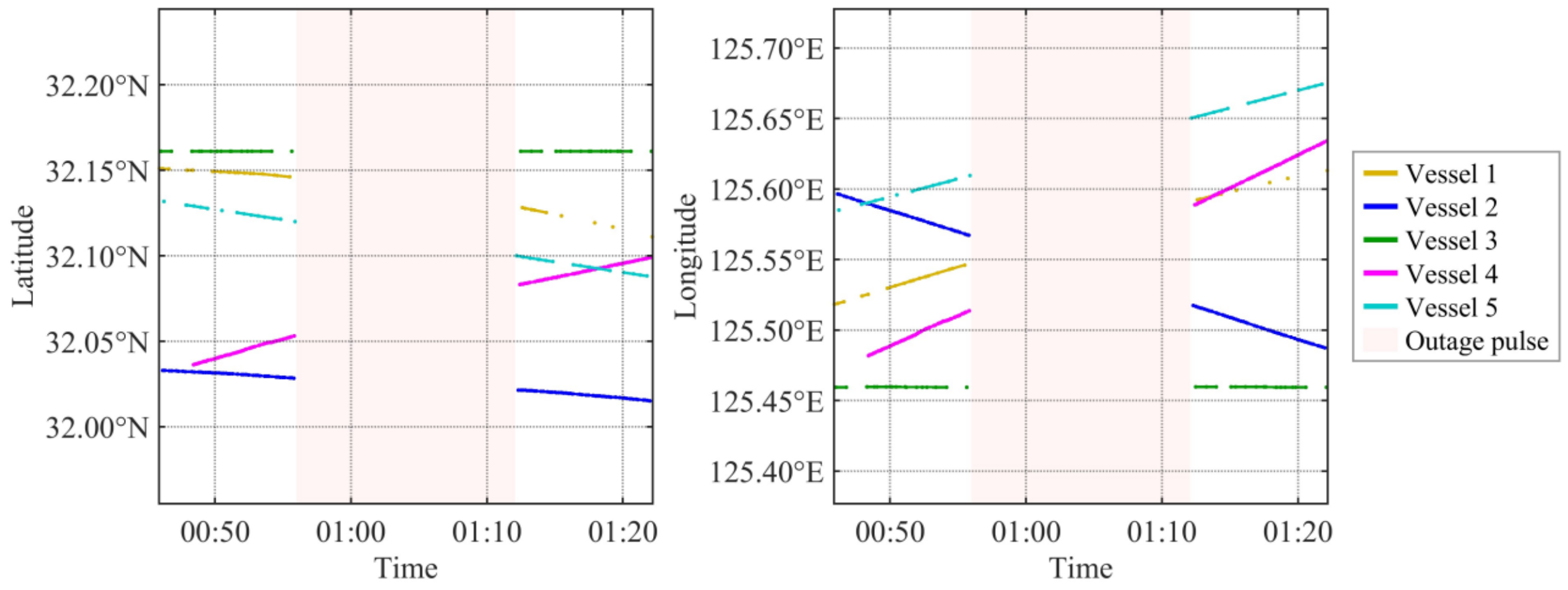

**FIGURE** 8 Representative jamming-consistent event on 26 December 2024: Time–Latitude and Time–Longitude diagrams showing a reporting-outage pulse across five vessels within approximately 10.0 km, lasting 977 s

The Time–Latitude and Time–Longitude diagrams show a single outage pulse from 00:55:53 to 01:12:10, lasting 977 s, during which no position reports were received from any of the five affected vessels. The aligned onset and recovery of the reporting loss across the affected vessels suggest a common reporting-continuity disruption rather than independent vessel-level failures.

Figure 9 provides the corresponding spatial and feed-level context. During the outage pulse, AIS transmissions from 356 surrounding vessels within approximately 150 km of the affected area continued to be received, and the whole-feed reception level remained near baseline while messages from the five affected vessels dropped to zero. This pattern supports interpretation as a localized reporting-outage event rather than a persistent receiver-coverage gap or a complete AIS feed outage.

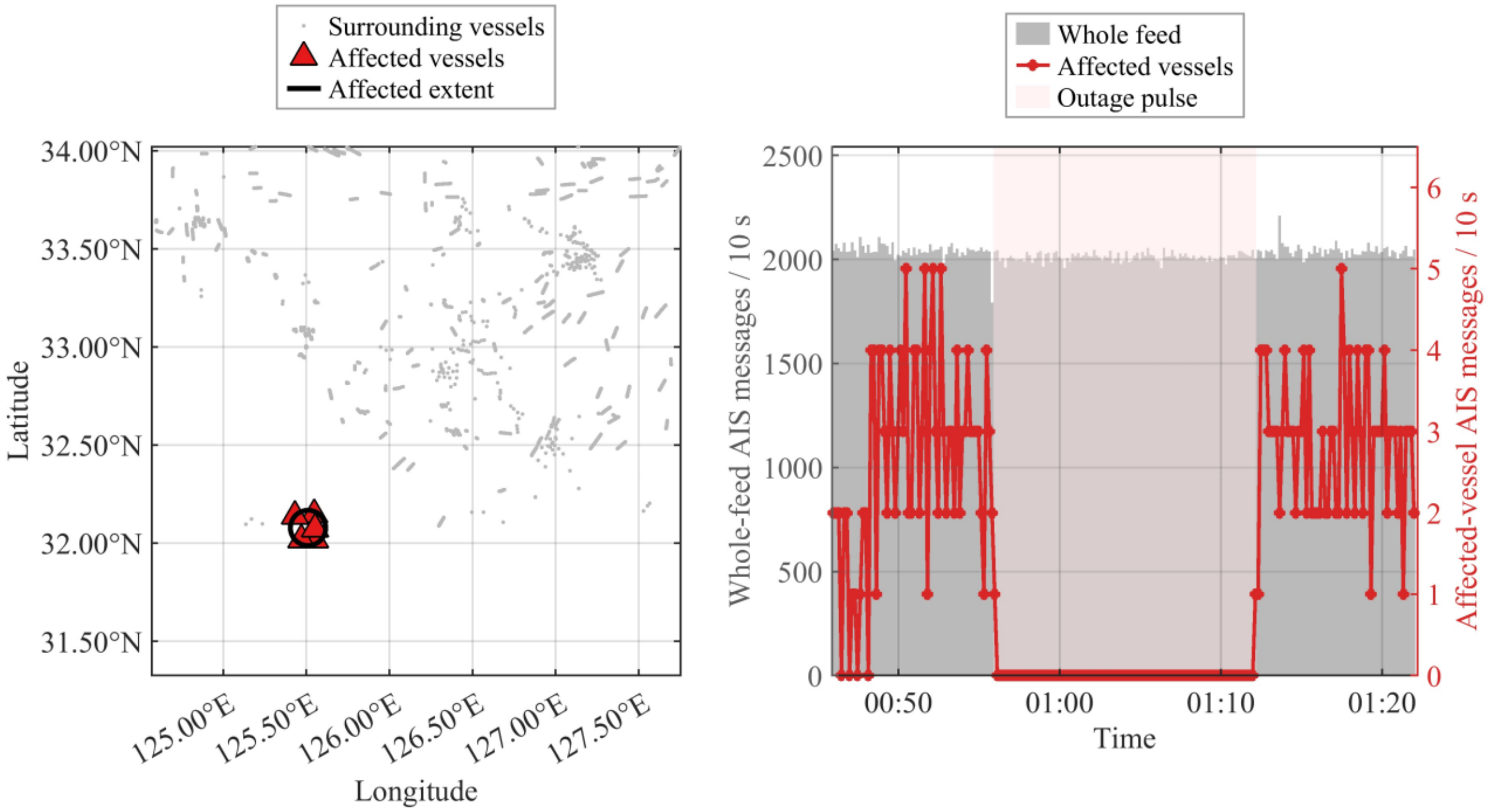

**FIGURE** 9 Representative jamming-consistent event: (left) spatial distribution during the outage pulse, showing active surrounding vessels and the affected-vessel extent (R = 10.0 km); (right) feed-activity panel showing that whole-feed reception remained near baseline while reports from the five affected vessels dropped to zero

Although AIS-only observations cannot independently confirm the physical radio-frequency (RF) source, the bounded outage timing, normal pre-event visibility, rapid post-event recovery, continued surrounding reception, and pre- and post-outage trajectory continuity make independent vessel-level transmission control or a simple receiver-coverage gap less likely. Therefore, this case is interpreted as a jamming-consistent localized reporting-outage event.

## 6 DISCUSSION

### 6.1 Classification reliability through hierarchical integrity separation

One of the primary challenges in AIS-based GNSS interference monitoring is the prevalence of non-interference anomalies that visually resemble spoofing or jamming. As demonstrated in Section 5, a large fraction of initial anomaly cues originate from communication-integrity issues and vessel-specific sensor faults rather than from externally induced GNSS disturbances.

The hierarchical separation of communication integrity assessment from subsequent spatiotemporal analysis is central to this classification reliability. By removing MMSI duplication and stale-data retransmissions prior to clustering, the framework prevents communication-layer artifacts from propagating into kinematic or multi-vessel analyses. This design choice is particularly important, as these artifacts can otherwise generate dense spatiotemporal structures that are indistinguishable from GNSS-RFI-consistent regional events when viewed solely at the trajectory level.

Similarly, the classification of single-vessel anomalies into persistent and transient sensor-integrity artifacts ensures that vessel-specific faults are not misinterpreted as regional interference events. The results in Section 5.2 illustrate that even visually structured deviations can be reliably excluded from GNSS interference categories when temporal persistence and cross-vessel coherence are jointly considered.

### 6.2 Role of spatiotemporal coherence in GNSS-RFI-consistent event identification

The case studies in Section 5.3 highlight the central role of spatiotemporal coherence in distinguishing GNSS interference from isolated positioning anomalies. Spoofing-consistent events are characterized by coordinated trajectory deviations across multiple vessels within a confined spatial region and a narrow temporal window, whereas jamming-consistent events manifest as synchronized localized reporting outages across multiple vessels.

By operating on anomaly cues rather than raw trajectories, the proposed ST-DBSCAN-based clustering framework accommodates both spoofing and jamming events within a unified analytical structure. This flexibility allows the framework to detect jamming events that would otherwise remain invisible to purely kinematic approaches, while simultaneously avoiding over-reliance on communication-layer heuristics.

Multi-vessel spatiotemporal coherence is not treated as definitive evidence of GNSS interference by itself. Rather, it is evaluated after communication integrity assessment and vessel-specific sensor-integrity separation. Nevertheless, shared receiver, network, or infrastructure-side effects cannot be fully excluded from AIS observations alone. Therefore, the detected regional events are interpreted as GNSS-RFI-consistent monitoring outputs rather than signal-level source confirmation.

### 6.3 Validation challenges and multi-evidence framework

A fundamental challenge in GNSS interference detection research is the scarcity of ground truth data for real-world events. Unlike controlled experiments or simulated datasets, operational maritime GNSS interference presents unique validation difficulties due to its covert nature, transient occurrence, and the absence of systematic reporting infrastructure. Deliberate spoofing and jamming are typically conducted without acknowledgment, while vessel operators may not report incidents due to operational pressures, security concerns, or lack of awareness. In the AIS-based maritime monitoring context considered here, this limitation means that independent event-level validation data, such as official incident reports, synchronized RF measurements, or verified per-event labels, are generally unavailable for the detected events. Nevertheless, period-level contextual evidence exists for the present study: the evaluation period overlaps with officially announced GPS jamming incidents affecting vessels in the West Sea in November 2024, with similar incidents reported earlier in 2024 and formally raised with the ITU, ICAO, and IMO (The Korea Times, 2024). This supports the premise that genuine GNSS RFI occurred in the study region during the analysis window, even though individual detected events cannot be matched one-to-one to specific reported incidents. Given the absence of such event-level labels, cluster-level precision, false detection rate, and missed detection rate cannot be directly computed as in a fully labeled dataset or controlled RF experiment. The detected events are therefore interpreted as GNSS-RFI-consistent monitoring outputs, supported by communication-artifact mitigation, vessel-specific anomaly separation, and representative case-level evidence.

This limitation is not unique to the proposed framework but represents a field-wide challenge. Among the 44 AIS anomaly detection studies surveyed by Wolsing et al. (2022), none reported validation against confirmed GNSS interference events. Recent studies have similarly relied on indirect evidence: Ferreira et al. (2022) employed domain expert review for validation, while Liang et al. (2024) used synthetic spoofing injection for testing, explicitly acknowledging the impossibility of obtaining ground truth for real-world interference.

Future validation efforts should prioritize establishing collaborative monitoring networks with maritime authorities, conducting controlled interference experiments in designated zones, and integrating AIS data with complementary sensing modalities such as Automatic Dependent Surveillance–Broadcast (ADS-B) and ground-based GNSS monitoring stations. Development of standardized reporting protocols, analogous to aviation's NOTAM system, would further enable systematic ground truth collection for the maritime domain.

### 6.4 Operational limitations

Several inherent limitations of the proposed framework should be acknowledged. Fixed thresholds used in communication integrity assessment may introduce ambiguity in boundary cases under heterogeneous network synchronization conditions. The effectiveness of kinematic anomaly cue extraction further depends on onboard GNSS sensor quality, as elevated noise in aging or low-grade equipment can affect the stability of IMM-based residuals. In addition, the framework relies on sufficient vessel density to establish spatiotemporal coherence and may exhibit reduced sensitivity to region-level GNSS interference in sparsely trafficked areas. Finally, while the classification logic distinguishes spoofing-consistent distortion events from jamming-consistent reporting-outage events at the phenomenological level, precise interference source localization and signal-level characterization remain beyond the scope of AIS-only analysis.

The framework is not intended to provide definitive attribution for isolated single-vessel anomalies. In particular, a single MMSI-overlap pattern may be difficult to distinguish from genuine MMSI duplication using AIS data alone. The proposed co-occurrence gate therefore prioritizes regional multi-vessel evidence and may conservatively exclude isolated overlap cases as communication-integrity artifacts. This design is appropriate for regional multi-vessel monitoring, but it limits sensitivity to single-vessel GNSS spoofing or low-density scenarios. Similarly, an isolated single-vessel reporting outage cannot be uniquely attributed to GNSS jamming using AIS observations alone, because similar gaps may result from intentional AIS silence, onboard receiver or transponder faults, local operational reporting control, or temporary reception loss. Therefore, isolated reporting gaps are retained only as low-level transmission-continuity cues and are not interpreted as jamming-consistent events unless they are supported by multi-vessel coherence and feed-alive reception context.

## 7 CONCLUSION

This study presented an AIS-based framework for maritime GNSS RFI monitoring by explicitly separating communication-integrity issues, vessel-specific sensor-integrity artifacts, and spatiotemporally coherent regional disturbances. By combining rule-based diagnostics with spatiotemporal clustering of anomaly cues, the framework mitigates communication-layer artifacts and vessel-specific irregularities while preserving sensitivity to both spoofing-consistent distortion events and jamming-consistent reporting-outage events.

Evaluation using a large-scale national AIS dataset demonstrated that the majority of apparent anomalies originate from non-interference sources, including communication artifacts and vessel-specific sensor faults. The results further showed that spatiotemporal coherence across multiple vessels provides a robust and interpretable basis for distinguishing GNSS-RFI-consistent regional events from visually similar but benign irregularities.

Future work will focus on extending the framework toward near-real-time monitoring, validating parameter robustness in diverse operational regions, and integrating complementary sensing modalities to support interference source characterization. The proposed approach offers a scalable and interpretable foundation for maritime GNSS interference monitoring using AIS data.

## ACKNOWLEDGMENTS

This research was supported by the Regional Innovation System & Education (RISE) program through the Chungbuk Regional Innovation System & Education Center, funded by the Ministry of Education (MOE) and Chungcheongbuk-do, Republic of Korea (2025-RISE-11-014-03). This research was also supported by the Korea Research Institute of Ships and Ocean Engineering through the Endowment Project "Digital Platform Development to Support Marine Digital Transformation," funded by the Ministry of Oceans and Fisheries (2520000291, PES5581).